\documentclass[
10pt,tightenlines, amsmath, amssymb, nofootinbib, prd,
superscriptaddress, showpacs, preprintnumbers]{revtex4}

\newcommand{\beq}{\begin{equation}}
\newcommand{\eeq}{\end{equation}}
\newcommand{\bea}{\begin{eqnarray}}
\newcommand{\eea}{\end{eqnarray}}
\newcommand{\bmat}{\begin{pmatrix}}
\newcommand{\emat}{\end{pmatrix}}

\def\pb{\overline{\psi}}
\def\gm{\gamma^{\mu}}
\def\d{\partial}
\def\<{\langle}
\def\>{\rangle}
\def\+{\dagger}
\def\Ht{H^{\bot}}
\def\dm{{\partial}_{\mu}}
\def\eps{\epsilon^{\mu \nu \lambda \sigma}}

\begin{document}
\title{  Anomalous Axion Interactions  and
Topological    Currents   in Dense Matter }
\affiliation{Department of Physics and Astronomy, University of
British Columbia, Vancouver, BC, Canada, V6T 1Z1}
\author{Max~A.~Metlitski}
\email{mmetlits@physics.ubc.ca} \affiliation{Department of Physics
and Astronomy, University of British Columbia, Vancouver, BC,
Canada, V6T 1Z1}
\author{Ariel~R.~Zhitnitsky}
\email{arz@physics.ubc.ca} \affiliation{Department of Physics and
Astronomy, University of British Columbia, Vancouver, BC, Canada,
V6T 1Z1}

\date{\today}
\begin{abstract}
Recently an effective Lagrangian for the interactions of photons,
Nambu-Goldstone bosons and superfluid phonons in dense quark
matter has been derived using anomaly matching arguments. In this
paper we illuminate the nature of certain anomalous terms in this
Lagrangian by an explicit microscopic calculation.
We also generalize the corresponding construction to introduce the
axion field. We derive an anomalous  axion effective Lagrangian
describing the interactions of axions with photons
   and superfluid phonons in the dense matter background. This
effective Lagrangian, among other things,
 implies that an axion current will be induced
in the presence of magnetic field. 
 We speculate that  this current may be
responsible for the explanation of neutron star kicks.
\end{abstract}

\maketitle
\section{Introduction}
It has been recently realized\cite{SZ} that some very unusual effects may take place
 in  dense matter $QCD$
in the presence of topological defects and/or external magnetic field.
 It is known that at large baryon density
 many global symmetries of $QCD$ are spontaneously
broken\cite{CFLRev,Alford}. In particular, it is expected that the
chiral symmetry will be spontaneously broken. This leads to
appearance of pseudo-scalar Goldstone bosons, which are generated
out of the vacuum by axial chiral currents. The spontaneous
breaking of global symmetries also leads to existence of
topological defects: domain walls and strings in dense
$QCD$\cite{SSZ,ForbesZ}. In a recent paper\cite{SZ} it was shown
that the effective Lagrangian for the interaction of Goldstone
bosons with the electromagnetic field in the presence of chemical
potential $\mu$ contains terms, which imply that topological
defects such as axial vortices and domain walls in dense $QCD$
carry electric current and magnetization respectively. The
corresponding effective Lagrangian was derived in \cite{SZ} in a
formal way
 by treating the fermion chemical potential as the zeroth
component of a fictitious vector gauge field $V_{\mu}$ and
appealing to chiral anomalies induced by $V_{\mu}$.

The main goal of this paper is twofold. First, we observe that one
of the terms in the effective Lagrangian derived in \cite{SZ}
implies that flux tubes in dense quark matter carry an axial
current. We
investigate the microscopic origin of this current. 
Second, we generalize the derivation\cite{SZ} to include the axion
field which, if exists, may play an important role in astrophysics
and cosmology.

This paper is organized as follows.  In section II, we will be
studying axial currents on flux tubes in dense matter. We will
confirm the existence of these currents in three ways: a) by
appealing to the effective Lagrangian derived in \cite{SZ}, b)
directly from the chiral anomaly due to the fictitious field
$V_{\mu}$, c) from a microscopic calculation. 
These three methods agree.

 In Section III, we  derive  the
anomalous  axion effective Lagrangian using previously established methods.
The corresponding effective Lagrangian  describes  the interactions of axions with photons
   and superfluid phonons in the dense matter background.   We speculate regarding some
phenomenological implications of the obtained results.

\newpage
\section{Axial Current on Magnetic Flux Tubes}
\subsection{Short Overview}
The appearance of fermion zero modes on topological defects is a
mathematically rich phenomenon with links to index theory of
elliptic operators and quantum anomalies, see original
papers\cite{vortex} and review\cite{GordonRev}. In particular, it
is known\cite{vortex, GordonRev, Aharonov} that flux tubes in
$2+1$ dimensional $QED$ possess fermion zero modes, whose
existence can be proved using trace identities and $2$-dimensional
Euclidean chiral anomaly equations. In the context of $2+1$
dimensional $QED$ these zero modes lead to a degeneracy of the
flux tube ground state and induction of a Chern-Simons term in the
effective bosonic action. However, when we consider massless $QED$
in $3+1$ dimensions, the zero modes are free to move up or down
the flux tube, depending on their chirality and polarization. We
show in this paper, that for massless Dirac fermions in $3+1$
dimensions at finite fermion chemical potential $\mu$ and
temperature $T$ in the background of a magnetic flux tube, the
zero modes  generate an axial current of $J = \frac{e \mu}{2
\pi^2} \Phi$ along the flux tube, where $e$ is the fermion charge
and $\Phi$ is the total magnetic flux. The current is topological
in nature as it depends only on the total flux and not on the
particular details of distribution of the magnetic field. This
result is ${\it exact}$ (at least if the magnetic flux does not
fluctuate) as the contribution of all, but the zero modes to the
axial current vanishes.

As already noted, the appearance of axial current on flux tubes
can be derived by using a somewhat different aspect of chiral
anomalies. Here the anomaly resides directly in $3+1$ dimensions
and appears when one thinks of the fermion chemical potential as
the zeroth component of a fictitious vector gauge field $V_{\mu}$.
Such a gauge field, as well as the ordinary electromagnetic field
$A_{\mu}$, contributes to the anomalous non-conservation of axial
current. If the chiral symmetry is spontaneously broken in the
system under consideration (say in dense $QCD$) and a
corresponding Goldstone boson $\eta$ appears, one requires the
effective Lagrangian for $\eta$ to reproduce the axial current
non-conservation. As will be shown below, the effective Lagrangian
for $\eta$, originally derived in \cite{SZ}, then implies the
appearance of axial current on flux tubes, which agrees with the
microscopic zero mode result. We also show that, alternatively,
one can obtain this result starting directly from the axial
current non-conservation equation, without appealing to the
effective Lagrangian for the $\eta$ boson. This later method, as
well as the microscopic derivation, imply that axial current on
flux tubes appears even if the chiral symmetry is not broken
spontaneously.

We note that although the presence of zero modes on flux tubes in
$QED$ and the symmetry of higher energy states, which makes the
microscopic calculations of this paper possible, have been
previously known\cite{Aharonov}, the problem of calculation of
axial current on flux tubes at finite chemical potential and
temperature, to our knowledge, has not been considered before.
Moreover, the computation of the axial current at finite chemical
potential using fictitious anomalies in $3+1$ dimensions is
certainly a new trick. This work confirms the validity of this
trick by an explicit microscopic calculation, thus, supporting the
original derivation of the anomalous effective
Lagrangian\cite{SZ}.
We note the similarity of computation of axial current on flux
tubes in $QED$ undertaken in the present paper to the recent
computation of electric current on cosmic strings at finite
chemical potential\cite{StringsM}, which was also motivated by the
anomalous terms in the effective Lagrangian for Goldstones in
dense $QCD$. Both calculations rely on the idea of zero modes with
fixed quantum numbers running along a $2$-dimensional topological
object uniform in the $3$rd direction and yield similar results
for the current.

\subsection{Anomalous Terms}
Here we briefly review the anomaly based arguments of
\cite{SZ}, which indicate that when an axial-like symmetry is
spontaneously broken at finite fermion chemical potential, the
effective Lagrangian for the corresponding Goldstone mode receives
a correction, topological in nature. We show that in the presence
of a background magnetic field, this correction leads to the
appearance of axial current on magnetic flux tubes.

Consider $QCD$ at large baryon chemical potential. As is well
known, such a system spontaneously breaks various global
symmetries of $QCD$\cite{CFLRev,Alford}, leading to the existence
of Goldstone bosons, whose number and form depends on the number
of light (massless) quark flavours $N_f$. Here, we consider a
general neutral Goldstone boson $\eta$,\footnote{This need not be
{\it the} ``$\eta$" boson of $QCD$.} with the following
transformation properties under one of the diagonal axial
symmetries of $QCD$\footnote{As was shown in\cite{SSZ}, all
instanton effects are suppressed in high-density $QCD$ and at very
large baryon chemical potential the explicit breaking of the
formally anomalous symmetry $U(1)_A$ becomes very weak.}: \beq
\label{transf}\psi_a \rightarrow e^{i Q_a \theta \gamma^5} \psi_a,
\;\;\; \eta \rightarrow \eta + \theta \eeq Here $Q_a$ denotes the
flavor content of the Goldstone boson, $a = 1 .. N_f$, and $\eta$
is created out of the vacuum by the current, \beq \label{Jf}
j^{\mu} =\sum_a Q_a \bar{\psi}_a \gamma^{\mu} \gamma^5 \psi_a\eeq

As is well known, it is useful to represent quark chemical
potentials as the zeroth components of a fictitious vector gauge
field $V_{\mu} = (1,\vec{0})$. Then the coupling of quarks to
$V_{\mu}$ and to the usual electromagnetic gauge field $A_{\mu}$
takes the form: \beq {\cal L} = \sum_a (\mu_a V_{\mu} - e_a
A_{\mu}) \pb_a \gm \psi_a \eeq where $\mu_a$ and $e_a$ are quark
chemical potentials and electromagnetic charges respectively. The
anomaly equation for the current $j^{\mu}$ in the background of
fields $V_{\mu}$ and $A_{\mu}$ takes the form: \beq \label{anom}
\dm j^{\mu} =  \eps (C_{\eta A A} \,F_{\mu \nu} F_{\lambda \sigma}
+ C_{\eta A V} \,V_{\mu \nu} F_{\lambda \sigma} + C_{\eta V V}
\,V_{\mu \nu} V_{\lambda \sigma})\eeq where the field tensors
$F_{\lambda \sigma}, ~ V_{\lambda \sigma}$ are defined as,
$F_{\lambda \sigma}\equiv \d_\lambda A_{\sigma} -\d_\sigma
A_{\lambda}, ~~ V_{\lambda \sigma}\equiv \d_\lambda V_{\sigma}
-\d_\sigma V_{\lambda}$, and the coefficients, \beq C_{\eta AA} =
- N_c \sum_a \frac{e_a^2 Q_a}{16 \pi^2},\;\; C_{\eta AV} = N_c
\sum_a \frac{e_a \mu_a Q_a}{8 \pi^2},\;\; C_{\eta VV} = -N_c\sum_a
\frac{\mu_a^2 Q_a}{16 \pi^2} \eeq The anomalous current
non-conservation (\ref{anom}) must be reproduced in the effective
Lagrangian for the neutral Goldstone boson $\eta$. Thus, as was
shown in\cite{SZ}, the effective Lagrangian for $\eta$ pics up the
following anomalous term describing its interaction with the
fields $A_{\mu}$ and $V_{\mu}$: \beq \label{anL1} L_{\eta} =
L^0_{\eta} + 2 \dm \eta \eps (C_{\eta A A} A_{\nu} F_{\lambda
\sigma} + C_{\eta A V} V_{\nu} F_{\lambda \sigma} + C_{\eta V V}
V_{\nu} V_{\lambda \sigma})\eeq Here $L^0_{\eta}$ is the standard,
non-anomalous part of the effective Lagrangian for $\eta$, which
transforms as $L^0 \rightarrow L^0 - j^{\mu} \dm \theta$ under
(\ref{transf}). We now restore the fictitious field $V_{\mu}$ to
its true value $V_{\mu} = (1, \vec{0})$. Then the last term in
(\ref{anL1}) vanishes\footnote{However, as shown in \cite{SZ} this
term can become important if the quark matter is rotating and/or
superfluid vortices appear.}, and we are left with, \beq
\label{anL2} L_{\eta}= L^{0}_{\eta} - C_{\eta A A} \, \eta \, \eps
F_{\mu \nu} F_{\lambda \sigma} + 4 C_{\eta A V} \nabla \eta \cdot
\vec{B} \eeq where $\vec{B}$ is the magnetic field. The first
anomalous term in eq. (\ref{anL2}) describes the usual anomalous
decay of a Goldstone boson to two photons, and is absent on the
classical level, if there is no background electric field present.
We now concentrate on the second anomalous term in (\ref{anL2}),
which does not occur in vacuum (at $\mu = 0$): \beq \label{anL3}
L_{\eta} = L^0_{\eta} + 4 C_{\eta A V} \nabla \eta \cdot \vec{B}
\eeq One effect of the new anomaly term originally discussed in
\cite{SZ} is the magnetization of the domain walls formed by the
$\eta$ field (such domain walls are possible, say, if $\eta$ is
associated with spontaneous breaking of the $U(1)_A$ symmetry,
which is also explicitly slightly broken by instantons\cite{SSZ}).
Here we discuss a different consequence of this term. Let's vary
the action obtained from Lagrangian (\ref{anL3}), with respect to
$\eta \rightarrow \eta + \theta$, to derive the classical
equations of motion. By construction, $L^0 \rightarrow L^0 -
j^{\mu} \dm \theta$, hence, \beq \delta S = \int d^4 x (- j^{\mu}
\dm \theta + 4 C_{\eta A V} B^i \d_i \theta) = \int d^4x \, \dm
j^{\mu} \, \theta + \int dt \int_{\d R} dS_i (-j^i + 4 C_{\eta A
V} B^i) \theta \eeq Here the surface integral is over the boundary
of the region $R$ where our dense matter is realized. So, as
$\nabla \cdot \vec{B} = 0$, the anomalous term does not contribute
to the equation of motion $\dm j^{\mu} = 0$. However, if we do not
restrict $\theta$ to vanish on the boundary, we also obtain a
boundary condition, \beq \label{bc} \vec{j} \cdot d\vec{S} = 4
C_{\eta A V} \vec{B} \cdot d\vec{S} \eeq Now, in the steady state
situation, there is no build up of axial charge, and we have
$\nabla \cdot \vec{j} = 0$. Hence for any cross-section $S$ of the
region $R$ let $S_b$ be the part of $\d R$ such that $\d S = \d
S_b$. Then, \beq \label{curr1} \int_S d\vec{S} \cdot \vec{j} =
\int_{S_b} d\vec{S} \cdot \vec{j} = 4 C_{\eta A V} \int_{S_b}
d\vec{S} \cdot \vec{B} = 4 C_{\eta A V} \int_{S} d\vec{S} \cdot
\vec{B} = N_c \sum_a \frac{e_a \mu_a Q_a}{2 \pi^2} \Phi \eeq where
$\Phi$ is the total magnetic flux through the cross-section $S$.
So we see that the anomalous term in eq. (\ref{anL3}) implies the
existence of an axial current flowing through the dense matter
which is proportional to the magnetic flux.

At this point we make the following important remark regarding the
formula (\ref{curr1}): the final expression for the current does
not depend on the specific properties of the pseudo-Goldstone
boson $\eta$, such as its coupling constant $f_{\eta}$. This is
not due to our choice of units, and this is not a typo, so it
leads us to weaken our assumption of spontaneous chiral symmetry
breaking and existence of the $\eta$ Goldstone.

Indeed, the result (\ref{curr1}) can be derived in the following
way for a generic system of massless fermions at finite chemical
potential in the magnetic field, without appealing to the
effective Lagrangian for the $\eta$ Goldstone boson. Let's return
to the anomaly equation (\ref{anom}). We can think of the
fictitious field $V_0$, as taking the value $``1"$ inside the
region $R$ where the quark matter is realized and $``0"$ outside
(if we have a quark star, this is actually rather close to
reality, since the interface between quark matter and vacuum is
very narrow). Then, again, if no electric field is present, and if
the axial charge density is time independent, eq. (\ref{anom})
takes the form, \beq \label{bound} \nabla \cdot \vec{j} = 4 \,
C_{\eta A V} \, \nabla \cdot (V_0 \vec{B}) \eeq The right hand
side of eq. (\ref{bound}) vanishes both inside and outside $R$,
yielding $\nabla \cdot \vec{j} = 0$. However, integrating
(\ref{bound}) over a small Gaussian pillbox on the boundary of
$R$, and recalling that in the vacuum outside $R$, $\vec{j} = 0$,
we obtain on the inner boundary of $R$, \beq \vec{j} \cdot
d\vec{S} = 4 C_{\eta A V} \vec{B} \cdot d\vec{S} \eeq This is the
same result (\ref{bc}) that we obtained by minimizing the action
for the Goldstone $\eta$. From this, we readily obtain the
expression (\ref{curr1}) for the total current across any
cross-section of $R$.

The last derivation does not use anywhere the spontaneous breaking
of chiral symmetry and the existence of the $\eta$ Goldstone, and
relies solely on chiral anomalies. Thus, when the dynamics of our
problem are such that a Goldstone boson $\eta$ exists, the
appearance of axial current on flux tubes can be extracted from
the effective Lagrangian (\ref{anL3}) for the $\eta$ field.
However, the existence of such axial current does not depend on
spontaneous chiral symmetry breaking, but rather on the mere
existence of chiral symmetry. This observation is confirmed
microscopically in the next subsection, where the result
(\ref{curr1}) is reproduced in a simple $QED$-like system.

\subsection{Microscopic Arguments}
We will show in this section that the appearance of current on
magnetic flux tubes at finite chemical potential derived in the
previous section using the anomalous effective Lagrangian for
$\eta$, can be understood very simply microscopically within the
following model. As we already saw at the end of the previous
section, the existence of axial currents on flux tubes does not
rely on spontaneous chiral symmetry breaking.  Thus, we choose to
ignore all the strong interaction effects leading to formation of
the Goldstone field $\eta$, and consider
 only the following
Lagrangian, \beq {\cal L} = \pb \, i (\d_{\mu} + i e A_{\mu})\gm
\psi - m \pb \psi + \mu \pb \gamma^0 \psi . \eeq which describes
the interactions of a single light quark $\psi$ of mass $m$ with a
background electromagnetic field $A_{\mu}$, at finite baryon
chemical potential $\mu$. Hence, the discussion in this section
actually applies to any $QED$-like system at finite chemical
potential.

We are interested in the case of magnetic flux tubes, i.e.
$A_{\mu}$ is static and the magnetic field $\vec{B} = \nabla
\times \vec{A} = B(x,y) \hat{z}$ is uniform in the third direction
$z$. Our goal is to compute the total axial current $J^3_5 = \int
\, d^2 x \langle \pb \gamma^3 \gamma^5 \psi\rangle$ along the flux
tube. The Dirac Hamiltonian is, \beq H = -i (\d_i + i e A_i)
\gamma^0 \gamma^i + m\gamma^0 \eeq and the Dirac equation becomes,
\bea & &-H_R \psi_L + m \psi_R = E \psi_L
\\ & &m \psi_L + H_R \psi_R = E \psi_R \eea where we use the
conventions of Peskin and Schroeder and,  \beq H_R = (-i \d_i + e
A_i) \sigma^{i} \eeq So, $\psi_L = \frac{1}{m}(E-H_R)\psi_R$,
where \beq (H_R^2 + m^2) \psi_R = E^2 \psi_R \eeq Hence every
eigenstate $\psi_R$ of $H_R$ with eigenvalue $\epsilon$ generates
two solutions of the Dirac equation with energies $E = \pm
\sqrt{\epsilon^2 + m^2}$ and, \beq
\label{LR} \psi_{\pm} = {\left(\begin{array}{c} \psi_L\\
\psi_R\end{array} \right)}_{\pm} = (4 (m^2 +
\epsilon^2))^{-\frac{1}{4}} \left(\begin{array}{c}
\pm ((m^2 + \epsilon^2)^{\frac{1}{2}} \mp \epsilon)^{\frac{1}{2}} \psi_R\\
((m^2 + \epsilon^2)^{\frac{1}{2}} \pm \epsilon)^{\frac{1}{2}}
\psi_R
\end{array} \right) \eeq
Now we concentrate on the right sector $H_R \psi_R = \epsilon
\psi_R$. Due to invariance with respect to translation in $z$
direction, we go to momentum eigenstates $-i \d_3 \psi_R = p_3
\psi_R$ (we take the third direction to be periodic of length $L$,
and take the limit $L \rightarrow \infty$ at the end of the
calculation). In each momentum sector, the operator $H_R$ takes
form,
\bea H_R &=& p_3 \sigma^3 + \Ht \\
\Ht &=& (-i \d_a + e A_a) \sigma^a, \;\; a = 1,2. \eea

We note that $\{\sigma^3, \Ht\} = 0$. Hence, if $|\lambda\>$ is a
properly normalized eigenstate of $\Ht$ with eigenvalue $\lambda$
then $\sigma^3 |\lambda\>$ is a properly normalized eigenstate of
$\Ht$ with eigenvalue $-\lambda$. So, all eigenstates of $\Ht$
with non-zero eigenvalues are of form $|\lambda\>, \, |-\lambda\>
= \sigma^3 |\lambda\>$, where $\lambda > 0$. Also, $\sigma^3$ maps
zero eigenstates of $\Ht$ to zero eigenstates of $\Ht$ and hence
we can classify all zero modes of $\Ht$ by their eigenvalue under
$\sigma^3$.

The eigenstates of $H_R$ can now be expressed in terms of
eigenstates of $\Ht$. Clearly, $[H_R, {\Ht}^2] = 0$, so $H_R$ only
mixes states $|\lambda\>, |-\lambda\>$. For $\lambda > 0$, we
write, \beq \label{psiR} \psi_R = c_1 |\lambda\> + c_2 \sigma^3
|\lambda\>\eeq
where $c_1,c_2$ satisfy: \beq \left(\begin{array}{cc} \lambda & p_3 \\
p_3 & -\lambda \end{array}\right) \left(\begin{array}{c} c_1 \\
c_2\end{array}\right) = \epsilon \left(\begin{array}{c} c_1 \\
c_2\end{array}\right) \eeq Hence $\epsilon = \pm \sqrt{\lambda^2 +
p_3^2}$ and, \bea \label{cs} \left(\begin{array}{c} c_1\\
c_2\end{array}\right)_{\pm} = (4(\lambda^2 +
p_3^2))^{-\frac{1}{4}} \left(\begin{array}{c} \pm
sgn(p_3) ((\lambda^2 + p_3^2)^{\frac{1}{2}} \pm \lambda)^{\frac{1}{2}} \\
((\lambda^2 + p_3^2)^{\frac{1}{2}} \mp
\lambda)^{\frac{1}{2}}\end{array}\right) \eea So each eigenstate
of $\Ht$ with an eigenvalue $\lambda > 0$ generates two
eigenstates of $H_R$. \footnote{There are known
examples\cite{GordonRev}, such as fermion number appearing on
domain walls, when this is not strictly speaking true. Indeed,
some of the energy levels of $H_R$ are continuous rather than
discreet and the correspondence discussed above between the
eigenstates of $\Ht$ and $H_R$ need not preserve the density of
states. However, for the particular Hamiltonian $H_R$, it can be
shown that if $B(x) \rightarrow 0$ as $x \rightarrow \infty$
sufficiently fast, this problem does not arise.}

The zero modes of $\Ht$ are simultaneously eigenstates of $H_R$
with eigenvalue, \beq \epsilon = p_3 \sigma^3 \eeq Hence, when the
mass $m \rightarrow 0$, zero modes of $\Ht$ become gapless modes
of $H$ capable of travelling up or down the flux tube depending on
the sign of $\sigma^3$ and on the chirality. We will shortly see,
that at finite chemical potential, precisely these modes carry an
axial current along the flux tube.

The following quantity will be of particular importance to us: $N
= N_+ - N_-$, where $N_{+}$ and  $N_{-}$ are the numbers of zero
modes of $\Ht$ with $\sigma^3 = 1$ and $\sigma^3 = -1$
respectively. Observe, that if $|\lambda\>$ is a zero mode of
$\Ht$ with $|\lambda\> = \left(u,\;v\right)$ then, \bea {\cal D} v
= 0,\;\;\; {\cal D}^{\+} u = 0 \eea where, \beq {\cal D} = -i \d_1
- \d_2 + e (A_1 - i A_2) \eeq Hence $N_+ = dim(ker({\cal
D}^{\+}))$, $N_- = dim(ker({\cal D}))$, and, \beq N = Index(\Ht) =
N_+ - N_- =  dim(ker({\cal D}^{\dagger})) - dim(ker( {\cal D}))
\eeq

The index of the elliptic operator $\Ht$ has been computed in
numerous works\cite{vortex,GordonRev,Aharonov} using two types of
methods: i) complex analysis methods, ii) trace identities and
axial Euclidean anomaly in $2$ dimensions (this is particularly
interesting in the light of our using $4$ dimensional anomalies
above to derive axial currents on flux tube at finite $\mu$). The
zero modes have also been computed exactly for some simple
configurations of the gauge field\cite{Aharonov}. In general the
index is
given by: \bea Index(\Ht) = \frac{e \Phi}{2 \pi}\\
\Phi = \int d^2 x \, B^{3}(x) \eea Hence the index measures the
number of flux quanta through the $xy$ plane, which is in essence
a topological quantity.

Now let's proceed to compute the axial fermion current induced at
finite chemical potential $\mu$. For further generality, we also
include the effects of non-zero temperature $T$. The axial current
density in the third direction is given by, \beq j^3_5(x) = \pb(x)
\gamma^3 \gamma^5 \psi(x) = \psi_L^{\+} \sigma^3 \psi_L(x) +
\psi_R^{\+} \sigma^3 \psi_R(x)\eeq We wish to compute the
expectation value of the total current along the flux tube, $J^3_5
= \int d^2x \<j^3_5(x)\>$. At finite chemical potential and
temperature we have, \bea \<j^3_5(x)\> &=& \sum_{E} n(E)\,
\psi^{\dagger}_E(x) \gamma^0 \gamma^3 \gamma^5 \psi_E(x)
= \sum_{\epsilon} (n((\epsilon^2 + m^2)^{\frac{1}{2}}) +
n(-(\epsilon^2 + m^2)^{\frac{1}{2}}))\psi_{R\epsilon}^{\dagger}(x)
\sigma^3 \psi_{R\epsilon}(x)
\eea Here, $n(E) = \frac{sgn(E)}{e^{\beta(E-\mu)sgn(E)} + 1}$ is
the usual Fermi-Dirac distribution, $\psi_E$ are eigenstates of
$H$ with energy $E$, $\psi_{R\epsilon}$ are eigenstates of $H_R$
with eigenvalue $\epsilon$, and we've used eq. (\ref{LR}). The
explicit form of $\psi_{R\epsilon}$ in terms of eigenstates of
$\Ht$ implies, \bea \label{J35long}\<J^3_5\> &=&
\frac{1}{L}\sum_{p_3}\sum_{\lambda
>0}\sum_{s = \pm}(n((\lambda^2 + p_3^2 + m^2)^{\frac{1}{2}}) +
n(-(\lambda^2 + p_3^2 + m^2)^{\frac{1}{2}}))
\langle\psi^s_R(\lambda,p_3)|\sigma^3|\psi^s_R(\lambda,p_3)\rangle
+
\\ &+& \frac{1}{L}\sum_{p_3}\sum_{\lambda = 0} (n((p_3^2 +
m^2)^{\frac{1}{2}})+n(-(p_3^2 + m^2)^{\frac{1}{2}})) \langle
\lambda |\sigma^3|\lambda\rangle \eea
Here $\lambda >0 $ label eigenstates of $\Ht$, which generate
eigenstates $\psi^\pm_R(\lambda,p_3)$ of $H_R$ with momentum $p_3$
and eigenvalue $\epsilon_{\pm} = \pm \sqrt{\lambda^2 + p_3^2}$,
while $\lambda = 0$ label the zero modes of $\Ht$. Now, let's
evaluate the matrix element $\langle
\psi^s_R(\lambda,p_3)|\sigma^3|\psi^s_R(\lambda,p_3)\rangle$ for
$\lambda > 0$. Using eq. (\ref{psiR}) and dropping the subscripts
$\lambda$, $p_3$, $s$, we obtain, $\langle
\psi_R|\sigma^3|\psi_R\rangle= (|c_1|^2 +
|c_2|^2)\langle\lambda|\sigma^3|\lambda\rangle + (c_1^* c_2 + c_1
c_2^*)$. Noting, $\langle\lambda|\sigma^3|\lambda\rangle =
\langle\lambda|-\lambda\rangle = 0$ for $\lambda > 0$, and using
the explicit formula (\ref{cs}) for $c_1$, $c_2$, we obtain,
$\langle\psi_R|\sigma^3|\psi_R\rangle = s p_3(\lambda^2 +
{p_3}^2)^{-\frac{1}{2}}$. This matrix element is odd in both $p_3$
and $s$, hence the sum over all $\lambda > 0$ in eq.
(\ref{J35long}) vanishes , and only the zero modes of $\Ht$
contribute to $J^3_5$. The zero modes carry a definite value of
$\sigma_3$, so that $\langle \lambda|\sigma^3|\lambda\rangle =
\sigma^3$. Thus, we are left with,  \bea \label{J35} J^3_5 &=&
(N_+ - N_-) \frac{1}{L}\sum_{p_3}(n((p_3^2 + m^2)^\frac{1}{2}) +
n(-(p_3^2 + m^2)^\frac{1}{2}))  = \frac{e \Phi}{2 \pi} n_m(T, \mu)\\
n_m(T, \mu) &=& \int \frac{dp_3}{2\pi} (n((p_3^2 +
m^2)^\frac{1}{2}) + n(-(p_3^2 + m^2)^\frac{1}{2}))\eea Here,
$n_m(T,\mu)$ is just the number density of one-dimensional
two-component (Dirac) fermions of mass $m$ at finite temperature
$T$ and chemical potential $\mu$. Hence, our final result
(\ref{J35}) is topological in nature, since for each value of $T$
and $\mu$, it is sensitive only to the total magnetic flux and is
independent of the particular distribution of the magnetic field.

Several limits of the result (\ref{J35}) are noteworthy. First of
all, in the massless limit $m \rightarrow 0$, one has $n(T, \mu) =
\frac{\mu}{\pi}$ and, \beq \label{J3massless} J^3_5 = \frac{e
\mu}{2 \pi^2} \,\Phi \eeq 
If there are several species of quarks present, we can sum eq.
(\ref{J3massless}) over quark flavours and colours to obtain the
current $J^3$ of eq. (\ref{Jf}), which in the true dense $QCD$
creates the $\eta$ boson, \beq \label{J36}J^{3} = N_c \sum_a
\frac{e_a Q_a \mu_a}{2 \pi^2} \, \Phi \eeq This agrees with our
result (\ref{curr1}) of the previous subsection, where we
explicitly used the fact $m = 0$ (and, hence, chiral symmetry) in
assuming that the axial current conservation is violated only by
anomalies. So, we see that the appearance of axial current on flux
tubes, which was derived somewhat mysteriously in the previous
section using the trick of fictitious chiral anomalies, is
microscopically due to fermion zero modes. Our microscopic
approach supports the validity of the fictitious chiral anomaly
trick and serves as a check of the anomalous effective Lagrangian
derived in \cite{SZ}.

Let us note that the result (\ref{J3massless}) is also independent
of temperature for $m = 0$, which is a quite natural feature of a
truly topological phenomenon. More explicitly, this fact is due to
the special property of massless one-dimensional fermions, namely,
their density at finite chemical potential is temperature
independent.

For arbitrary mass $m \neq 0$, the density of one-dimension
fermions $n(T, \mu)$ is generally temperature dependent, so for
simplicity we consider the limit $T = 0$. Then, $n(0, \mu) =
\sqrt{\mu^2 - m^2}/{\pi}$ and, \beq \label{currm} J^3_5 = \frac{e
\sqrt{\mu^2 - m^2}}{2 \pi^2} \, \Phi \eeq It is instructive to
take the non-relativistic limit of eq. (\ref{currm}). Writing,
$\mu = m + \mu_{nr}$, where the non-relativistic chemical
potential $\mu_{nr} \ll m$, \beq J^3_5 \approx \frac{e \sqrt{2 m
\mu_{nr}}}{2 \pi^2} \Phi \eeq In the non-relativistic setting,
$J^3_5$ is just the spin $S^3$, and for the case of uniform
magnetic field, our result stems from the familiar fact that all
Landau levels are doubly degenerate with respect to spin, except
the lowest Landau level. It is amazing that this simple fact has
such deep connections to chiral anomalies in $2$ and $4$
dimensions.

\section{axion}
This section is devoted to the derivation
of the anomalous effective lagrangian including the axion field using some
previously developed methods.
 While   the axion is considered to be one of the best dark matter candidates
 (see original papers\cite{PQ}-\cite{KSVZ}, and reviews \cite{rev}),
 it has not been discovered yet. We derive novel low energy terms, which
 describe the interaction of  the axion field with other light particles: photons  and superfluid phonons
 in dense matter background.
  These terms may lead to phenomenologically important
 effects related to the axion astrophysics, which were not  discussed previously.


  We define the $\theta$ term in the fundamental $QCD$ lagrangian in the standard way,
$L_{\theta}=\frac{g^2\theta}{32\pi^{2}}\tilde{G}^{\mu\nu a}G_{\mu\nu a}$.
 The existence of the $\theta$ term implies a violation of P,
CP and T symmetries. However, there is no experimental evidence
for  P or CP violation in strong interactions. For example, CP
violation in QCD would induce electric dipole moments of strongly
interacting particles and there are stringent experimental limits
on those quantities. Thus the absence of CP violating effects in
$QCD$ indicates a very small value for the parameter $\theta$: why
is $\theta$ so small?

 The most elegant resolution was proposed by Peccei and Quinn who assumed that
the strong interactions Lagrangian has a global $U(1)_{PQ}$
chiral symmetry\cite{PQ}.
 Weinberg and Wilczek\cite{WW}
 analysed the consequences of the Peccei-Quinn symmetry
and  noticed that the  spontaneous breaking
of a global chiral symmetry $U(1)_{PQ}$
  leads to a light pseudoscalar
pseudo-Goldstone boson, called an axion, that will interact with
topological charge density,  $\frac{g^2
}{32\pi^{2}}\tilde{G}^{\mu\nu a}G_{\mu\nu a}$. In
papers\cite{DFSZ,KSVZ} two different types of the  invisible axion
models were suggested where it was demonstrated that the strong CP
problem in $QCD$ can be successfully solved with arbitrarily weak
axion coupling constant.

The only information which is relevant
for us in what follows is    the  transformation
properties of quarks under $U(1)_{PQ}$ chiral symmetry,
 \beq
\label{pq} \psi_a \rightarrow e^{i\alpha Q^{PQ}_a   \gamma^5}
\psi_a, \eeq where $Q^{PQ}_a$ is the $PQ$ charge for quark species
$\psi_a$. We should note  that leptons and Higgs bosons  have also
nontrivial transformation properties under $U(1)_{PQ}$  symmetry,
however this part  is not essential   for the present paper.
  We note that the axion interacts with gluons
through  a triangle diagram,
$L_{agg}=\frac{a}{f_a}\frac{g^2 }{32\pi^{2}}\tilde{G}^{\mu\nu a}G_{\mu\nu a}$.
It also interacts with photons
$L_{a\gamma\gamma}\sim g_{a\gamma\gamma} a{\tilde{F}^{\mu\nu }}F_{\mu\nu }$
 with a coupling constant $g_{a\gamma\gamma}\sim 1/f_a\sum_aQ^{PQ}_a e_a^2  $
  expressed in terms of electric $e_a$
 and $PQ$ charges $Q^{PQ}_a$ of all quarks and leptons. In what follows, we identify the
 axion field with the dimensionless phase $\alpha(x)$ (\ref{pq}). Physical, dimensional field
 $a\sim f_a\alpha$.
\subsection {Anomalous Axion Lagrangian}
 The key observation here is as follows. The relevant for this work transformation properties
 of quarks under chiral rotations (\ref{transf})  and under $PQ$ rotations (\ref{pq}) are
 very similar, $Q_a\rightarrow Q^{PQ}_a$.  Therefore, we can literally follow our previous calculations
 (Section II) in order to derive
 the anomalous effective lagrangian  for the axion field  (which replaces the Goldstone field $\eta$)
 in the presence of chemical potential $\mu$. We use the same trick\cite{SZ} in this derivation
 by representing quark chemical potential as the zeroth component of a fictitious field $V_{\mu}$.
The result of this calculation is almost identical to
(\ref{anL1}), \beq \label{axion_eff} L_{\alpha} = L^0_{\alpha} + 2
\dm \alpha \eps (C_{\alpha \gamma\gamma} A_{\nu} F_{\lambda
\sigma} + C_{\alpha \gamma V}V_{\nu} F_{\lambda \sigma} +
C_{\alpha V V} V_{\nu} V_{\lambda \sigma}) \eeq where $
L^{0}_{\alpha} $ describes all non-anomalous terms including the
axion kinetic term $f_a^2(\d_{\mu}\alpha)^2$, as well as different
interaction terms of the axion with quarks and leptons.
Coefficients $ C_{\alpha \gamma\gamma}, ~~C_{\alpha \gamma
V},~~C_{\alpha VV}$ can be easily extracted from the calculation
of the triangle diagram and are given by,
 \beq
 \label{axion1}
 C_{\alpha \gamma\gamma} = -
  \sum_a \frac{e_a^2 Q_a^{PQ}}{16 \pi^2},\;\; C_{\alpha \gamma V} =
\sum_a \frac{e_a \mu_a Q_a^{PQ}}{8 \pi^2},\;\; C_{\alpha VV} = -
\sum_a \frac{\mu_a^2 Q_a^{PQ}}{16 \pi^2} , \eeq where label  $a$
runs   over all species (colours) of particles with nonzero $PQ$
charges including quarks and leptons. The first term $\sim
C_{\alpha \gamma\gamma} \alpha {\tilde{F}^{\mu\nu }}F_{\mu\nu }$
in eq. (\ref{axion_eff}) is the well known interaction between the
axion and photons.\footnote{In the literature devoted to the
axion\cite{rev} one typically includes in $ C_{\alpha
\gamma\gamma} $ the effects due to the mixing of the axion with
Goldstone fields such as the pion. We ignore this mixing for all
qualitative discussions in what follows.} In particular, this term
describes the axion decay into two photons. It also describes the
axion$\leftrightarrow\gamma$ transitions in the presence of the
magnetic field $B$. The corresponding effect plays a crucial role
in most axion search experiments. We shall not discuss this term
in the present paper.

Two other terms are new, and as far as we know, these terms have
never been discussed in the literature. The last term in eq.
(\ref{axion_eff}) vanishes in the topologically trivial
background. However, in the presence of superfluid vortices
similar to the case discussed in \cite{SZ}, this term, among other
things, describes the axion$\leftrightarrow $superfluid phonon
transitions. It might be phenomenologically important in rotating
neutron stars.
\subsection{Possible Applications}
Let us  concentrate on the middle
 term in brackets in eq. (\ref{axion_eff}). We rewrite it
in terms of the physical fields in the following way,
 \beq
  \label{axion}
 L_{\alpha}=  4
C_{\alpha \gamma V}\vec{\nabla} \alpha \cdot \vec{B}
\eeq
This interaction  explicitly shows that a Peccei-Quinn   current
corresponding to the  Peccei-Quinn  symmetry will be  induced in
the presence of an external magnetic field. Indeed, one can
literally follow the derivation (\ref{curr1}) to get   the
following expression for the  Peccei-Quinn  current, \beq
\label{axion2} J^{PQ} = \int_S d\vec{S} \cdot \vec{j} =
  4 C_{\alpha \gamma V} \int_{S} d\vec{S} \cdot
\vec{B} = \sum_a \frac{e_a \mu_a Q_a^{PQ}}{2 \pi^2} \Phi \eeq
where $\Phi$ is the total magnetic flux through the cross-section
$S$, and we have neglected the lepton chemical potentials. So we
see that the anomalous term in eq. (\ref{axion}) implies the
existence of  a Peccei-Quinn current  (\ref{axion2}) flowing
through the dense matter, which is proportional to the magnetic
flux.

A few remarks are in order. First, as we noticed previously, the
result (\ref{axion2}) does not depend on $f_a$ similar to the
previous case (\ref{curr1}) when the result did not depend on
$f_{\eta}$. It may look very suspicious because one can make $f_a$
arbitrarily large, which corresponds to an arbitrarily small
interaction of the original fermions with the axion. However, the
point is that the current (\ref{axion2}) corresponds to the
equilibrium state in an infinitely large bulk of matter. The
relevant question
  for the present case is:  what is  the formation time for such a current?
 It is obvious that with  $f_a$ increasing, the formation time
also increases, such that there is no contradiction with eq.
(\ref{axion2}) being independent of $f_a$.

Our next remark is that  the  Peccei-Quinn  (\ref{axion2}) as well as
the axial current (\ref{curr1}) are  unique  due to their topological nature.
 Indeed, even in the strongly interacting theory the axial current (\ref{curr1})
  is persistent and non-dissipating. It means that even
in such an unfriendly environment as the dense quark/nuclear
matter in neutron stars the current does not dissipate  due to
re-scattering and can be effectively used to deliver information
across the
bulk of the star. 
Therefore, there is a unique opportunity here to use our
topological currents (\ref{curr1}), (\ref{axion2}) for delivering
the asymmetry produced in the bulk of the star to solve the
problem of neutron star kicks\cite{kick,kick1}.

 The problem can be explained as follows. As is known,
pulsars exhibit rapid proper motions  characterized by a mean
birth velocity of $450 \pm 90$ km/s. Their velocities range from
100 to 1600 km/s\cite{kick}, while their distribution leans toward
the high-velocity end, with about 15$\% $ of all pulsars having
speeds over 1000 km/s\cite{kick1}. Pulsars are born in supernova
explosions; therefore, it would be natural to look for an
explanation in the internal dynamics of the supernova. However,
three-dimensional numerical
 simulations\cite{kick2} show that even the most extreme asymmetric explosions do not produce pulsar velocities greater than 200 km/s. Therefore, a different explanation should be found.
The origin of these motions has been the subject of intense study
and several possible
 explanations have been proposed. Many of the suggested mechanisms are capable (``in principle") to produce the required asymmetry.
 Indeed, in the presence of an external magnetic field, the produced neutrinos
 are automatically  asymmetric with respect to the direction of $\vec{B}$.
 However, the main common problem suffered by most suggested mechanisms is the
 difficulty of delivering the produced asymmetry to the surface of the
star. Only in this  case, the asymmetry may result in producing
the proper motion of the entire star. To overcome the difficulty
with delivery of the produced asymmetry to the surface, some
proposals, for example,  are based on new particles (such as a
sterile neutrino), which could
 escape from the bulk of the neutron star and deliver the asymmetry to the surface, see e.g. \cite{kusenko,Kusenko:2004wv}.\footnote{
 It is interesting to note that the interaction (\ref{axion}) is a precise
realization of the idea\cite{Kusenko:2004wv} that the magnetic
field may be correlated with the momentum of a very weakly
interacting particle, which can easily escape the star (Majoron or
sterile neutrino as suggested in \cite{Kusenko:2004wv}). Dynamics,
more precisely, anomaly, does the job of correlating the magnetic
field with the weakly interacting  axion current. } Our main
observation here is that due to their topological nature, the
currents (\ref{curr1}), (\ref{axion2}) may be capable of
delivering the required asymmetry  (produced in the interior of
the star) to the surface without dissipation\cite{MSZ}.

\section{Conclusion. Future Directions.}

In this paper we have discussed the appearance of axial current on
magnetic flux tubes at finite fermion chemical potential using
several approaches. All of these approaches are weaved together by
chiral anomalies. Microscopically, the current can be understood
in terms of fermion zero modes on the flux tube. These fermion
zero modes are in a certain sense themselves due to anomaly in $2$
dimensional Euclidean field theory and have implications for the
$2+1$ dimensional $QED$. Thus, we see that our trick with
fictitious anomalies at finite chemical potential in $3+1$
dimensions, in a sense continues the propagation of anomaly from
$2$ to $3$ to $4$ dimensions. This is a common pattern in the
study of anomalies.

We anticipate  a number of different applications of the derived
anomalous effective low energy lagrangian for the Goldstone bosons
and axion in dense matter. Some of them were mentioned in the
original paper\cite{SZ}, others were mentioned in the present
text. One specific application, which we believe deserves further
study is
the explanation of neutron star kicks\cite{MSZ}. 
In addition, novel anomalous effective lagrangian including the
axion field might be quite important for analysis of a number of
astrophysical problems, which would be the subject of a future
work.

\section*{Acknowledgements}
We are thankful to D.~T.~Son for discussions and critical remarks.
We also would like to acknowledge useful discussions with G.~E.~Volovik
and P.~B.~Wiegmann. We would also like to thank the organizers of
the program ``QCD and Dense Matter: From Lattices to Stars" at the
Institute for Nuclear Theory, Seattle, where this work was
initiated. This work is   supported in part by the Natural Sciences
and Engineering Research Council of Canada.

\end{document}